# The Nature is Simple in Essence

## - Blueprint for Unification of Gravity and Electromagnetism -

**Viorel Drafta**

drafta_v@yahoo.com

**Abstract**

The Temporal Fluctuation Model proposed in present paper represents an attempt for a unitary vision for gravity, inertia and electromagnetism. We obtain the main results of General Relativity and Electromagnetism and the theory, until now, don't contain self-contradictions. As well, we obtain the "unified potential" which has terms that coupling gravitation with electromagnetism and who could be the basis for future technological applications.

## 1. Introduction

Despite of many attempts, gravity and electromagnetism remain two separate phenomenons, although we are convinced that it must be a link between them.

The papers [19] and [16] show that it is possible to imagine other theories and interpretation than General Relativity and that they can explain the known experimental facts too.

The experimental fact for below proposed model is that the variation of event's duration increases at approaching of gravitational sources.

Starting from this, we develop a model for gravitational and electromagnetic interactions obtaining the main results of General Relativity and Electromagnetism. As well, we obtain the "unified potential" which has terms that coupling gravitation with electromagnetism and who could be the basis for future technological applications.

## 2. The Temporal Fluctuations Model (TFM)

It is natural to suppose that for a unitary vision of the reality the elemental constituents should have the simplest properties and interacts in the simplest possible way. The richness of reality should derive from selforganisation of elemental constituents based on their properties and interactions.

The model of gravitational and electromagnetic interactions proposed below is trying to correspond at this desirability. The experimental fact on which is based this model is that duration of events increase at approaching of a source of gravity [19], [8], [9].

We presume that elemental constituents could have the following properties:
- Have a finite space-temporal extension, i.e. have a finite volume and a proper lifetime;
- Are penetrable;
- Every constituent modifies the lifetime of each constituent who passes through. The ensemble of constituents (from a finite volume) modifies in this way the lifetime of each constituent of ensemble. Accordingly, the duration of events, the birth and extinction rate of constituents is modified too;
- They have moved each other with a finite relative velocity – we name it light speed - c. This property is important for maintain the relativity of inertial referential systems, as we shall see later.

Based on above properties, we consider that elemental constituents could be named *temporal fluctuations* (shortly: *fluctuations*).

Also, we nominate *intensity* - I - the measure of fluctuation's capability to modify the lifetime of other fluctuations. The value of intensity decreases from middle to edge and we could define an average intensity $I_m$ for each fluctuation.

It is natural to suppose that average intensity is different from one fluctuation to other, and that for the whole ensemble of fluctuations from any arbitrary small volume, it is a distribution of average intensity. We suppose that lifetime of fluctuations have a similar distribution to average intensity.

It is natural to presume that in the whole space (with Minkowski metric) the average density of fluctuations is constant (and equal with $n_0$) and with isotropic distribution of velocities.

In a given volume, for a given average intensity and a given velocity distribution, result a value for average lifetime and the ensemble of fluctuations is in an equilibrium state. The vacuum is characterized by average of fluctuations density $n_0$, average fluctuation intensity $I_0$ and average fluctuation lifetime $\Delta t_0$.

If on modify the average intensity and/or velocity distribution, on modify the lifetime, birth and extinction rate, but $n_0$ remain the same although it can suffer transient variations during modification phase.

We'll demonstrate that gravitation is linked with variation of average intensity and electromagnetism with altered distribution of velocities, because their sources make local modification of fluctuation's status.

### 3. Static Gravity and Electrostatics in Temporal Fluctuations Model

#### 3.1 The Mass and Radius of Particles

Given a gravitational source of mass M and a test body (particle) of rest mass $m_0$ at distance R from M, the energy of test mass is [8]:

$$E = m_0 c^2 \left(1 - \frac{GM}{c^2 R}\right) \qquad (1)$$

This can be understood as decreasing of body's rest mass when the body enters in gravitational field:

$$m = m_0 \left(1 - \frac{GM}{c^2 R}\right) \qquad (2)$$

We obtain relation (2) by considering that the amount of mass decrease when it enters in gravitational field is:

$$m = \frac{m_0}{k} \qquad (3)$$

Where k is a factor that represent the amount of rest mass decrease.

We introduce (3) in Lagrange equation for a free particle:

$$\frac{d}{dt}\left(\frac{m\vec{v}}{\sqrt{1-\frac{v^2}{c^2}}}\right) = \nabla\left(-mc^2\sqrt{1-\frac{v^2}{c^2}}\right) \qquad (4)$$

Working with spherical symmetry, radial movements, stationary case and v<<c, result:

$$\frac{dk}{dR} = \frac{1}{c^2} a(R) \qquad (5)$$

Where $a(R) = -GM/R^2$ is radial acceleration. By integrating equation 5 with condition:

$R \rightarrow \infty \quad \Rightarrow k \rightarrow 1$

It result:

$$k = e^{\frac{GM}{c^2 R}} \cong 1 + \frac{GM}{c^2 R} \qquad (6)$$

Where the right side is for weak field approximation. From (3) result:

$$m = \frac{m_0}{e^{\frac{GM}{c^2 R}}} \qquad (7)$$

For weak field (R>>0) we obtain equation (2). Moreover, if we substitute (2) in expression of free particle's Lagrangean, result the Lagrangean for the particle in weak gravitational field.

From (2) result the mass modification when the body enters in weak gravitational field:

$$\Delta m = - m_0 \frac{GM}{c^2 R} \qquad (8)$$

Now we fall back on the proposed model. We consider that all fluctuations have the same average intensity. Firstly, we define the density of average intensity by following relation:

$$i = I_0 n_0 \qquad (9)$$

We presume that at border of a particle with radius $r_0$ the density of average intensity is $i(r_0)$ and outside of particle it decrease following the law:

$$i(R) = i_0\left(1 + \frac{q}{R}\right) = k \cdot i_0 \qquad (10)$$

Where $i_0$ is density of average intensity for vacuum and q is constant. An argument for relation (10) is given in Appendix. We'll see that k from equations (6) and (10) are the same. It follows that at particle's border:

$$i(r_0) = i_0\left(1 + \frac{q}{r_0}\right) \qquad (11)$$

Now, we presume that rest mass is linked with difference between $i(r_0)$ and average intensity's density of surroundings. For a particle outside the gravitational field, mass is defined by following relation:

$$m_0 = C_m [i(r_0) - i_0] \qquad (12)$$

Where $C_m$ is constant for test particle. Rest mass at distance R from gravitational source with mass M is:

$$m = C_m\left[i(r_0) - i_0\left(1 + \frac{Q}{R}\right)\right] \qquad (13)$$

Where Q is constant analogue to q.

Result that the decrease of rest mass is: $\Delta m = -C_m i_0 (Q/R)$. By comparing with (8), we can identify:

$$m_0 = C_m i_0 \qquad (14)$$

*And*

$$Q = GM/c^2 \qquad (15)$$

For fulfilling (14), it must that $q = r_0$.

Also, result that we can write:

$$r_0 = Gm_0/c^2$$

*And* (16)

$$R_0 = GM/c^2$$

Where $R_0$ is the radius of gravitational source. We see that $R_0$ and $r_0$ are half of Schwarzschild radius of particles. By rewriting, we obtain:

$$m_0 = \frac{c^2}{G} \cdot r_0$$

(17)

$$M = \frac{c^2}{G} \cdot R_0$$

We conclude that radius of particle determine its mass. By comparing (12) with (14) we can conclude that at particle's border the density of intensity is always $2i_0$ and the only difference between particles is due to their radius.

At this moment, we can't say if the value $2i_0$ is proper only for border or for all volume of particle.

For particle's mass in gravitational field:

$$m = \frac{c^2}{G} \cdot r_0 \cdot \left(1 - \frac{GM}{c^2 R}\right)$$

(18)

We conclude also that particle's radius shrink when the particle comes near to gravitational source:

$$r = r_0 \left(1 - \frac{GM}{c^2 R}\right) \cong \frac{r_0}{k}$$

(19)

Equation (18) can be generalized for dimensions of all bodies that enter in gravitational field.

When we define mass, we presume that $i(r_0)$ is greater than $i_0$. If we consider the possibility that $i(r_0)<i_0$ it follows that $i(r_0)=0$ which signify that there no exist any fluctuations, fact who is absurd. However, it is possible that, in certain conditions, as transient effect, particle behave as if it has a mass smaller than its proper mass [5], [20], [21].

### 3.2 Interaction of Temporal Fluctuations

We consider that, in a small region of space, using an orthogonal coordinate system Oxyz, the number of fluctuations with velocities having the orientation at angles ($\phi,\theta$) that pass in a given time is:

$$N = C \cdot \left[1 + \left(\frac{\Delta N}{N_0}\right) \cdot f(\varphi, \theta)\right] \quad \varphi \in [0, 2\pi]; \quad \theta \in \left[-\frac{\pi}{2}, \frac{\pi}{2}\right] \quad (20)$$

Where:
- $N_0$ is the total number of fluctuations passing through that region in a given time;
- ($\Delta N/N_0$) is the maximum asymmetry of number of fluctuations: it is an orientation ($\phi_0$, $\theta_0$) at which $N=C[1+(\Delta N/N_0)]$ and an orientation ($\phi_0-\pi$, $-\theta_0$) at which $N=C[1-(\Delta N/N_0)]$;
- $f(\phi, \theta)$ is the distribution of fluctuations by orientation.

We find the value of C by integration for all $\phi$ and $\theta$. It result:

$$N = \frac{N_0}{2\pi^2 \left[1 + \left(\frac{\Delta N}{N_0}\right) \cdot a(\varphi_0, \theta_0)\right]} \cdot \left[1 + \left(\frac{\Delta N}{N_0}\right) \cdot f(\varphi, \theta)\right] \quad \varphi \in [0, 2\pi]; \quad \theta \in \left[-\frac{\pi}{2}, \frac{\pi}{2}\right] \quad (21)$$

For the sake of simplicity, we consider that all fluctuations have the same average intensity $I_0$ and the same average lifetime $\Delta t$.

Now, we focus on a fluctuation "j" that in its lifetime is subject of interaction from the others fluctuations. The influence must depend on number of fluctuations that interact with our fluctuation in its lifetime, their intensities (by average intensity) and asymmetry of fluctuation's distribution. We write down this influence:

$$\Phi_j = \sum_i \left[ C \cdot I_0 \cdot \Delta t \cdot N_i \cdot \left( h(\varphi_i - \varphi_j, \theta_i - \theta_j) + \left(\frac{\Delta N}{N_0}\right) \cdot g(\varphi_i - \varphi_j, \theta_i - \theta_j) \right) \right] \quad (22)$$

Where summation is on all fluctuations, C is constant and [h+($\Delta N/N_0$)g] is the factor due to asymmetry of fluctuation's distribution. The shape of [h+($\Delta N/N_0$)g] was choose to be simplest possible and that for isotropy the factor become equal to h that represent the interactions between fluctuations in absence of anisotropy.

By averaging Φⱼ for all fluctuations from a given volume V, we could write:

$$\Phi = \frac{1}{V}\frac{1}{N_0}\sum_j\left[N_j\cdot\sum_i\left[C\cdot I_0\cdot\Delta t\cdot N_i\cdot\left(h(\mathbf{j}_i-\mathbf{j}_j,\mathbf{q}_i-\mathbf{q}_j)+\left(\frac{\Delta N}{N_0}\right)\cdot g(\mathbf{j}_i-\mathbf{j}_j,\mathbf{q}_i-\mathbf{q}_j)\right)\right]\right] \quad (23)$$

By writing in integral form with average of fluctuations density on obtain:

$$\Phi = \frac{1}{n_0}\int\left[n_j\cdot\int\left(C\cdot I_0\cdot\Delta t\cdot n_i\cdot\left(h(\mathbf{j}_i-\mathbf{j}_j,\mathbf{q}_i-\mathbf{q}_j)+\left(\frac{\Delta n}{n_0}\right)\cdot g(\mathbf{j}_i-\mathbf{j}_j,\mathbf{q}_i-\mathbf{q}_j)\right)\right)d\mathbf{j}_i\,d\mathbf{q}_i\right]d\mathbf{j}_j\,d\mathbf{q}_j$$

(24)

By integrating equation (24), neglecting the terms with superior powers of (Δn/n₀) and using relation (9) we obtain:

$$\Phi = C\cdot\Delta t\cdot i\cdot\left(\mathbf{a}(\mathbf{j}_0,\mathbf{q}_0)+\mathbf{b}(\mathbf{j}_0,\mathbf{q}_0)\cdot\left(\frac{\Delta n}{n_0}\right)\right) \quad (25)$$

Where α(ϕ₀, θ₀) and β(ϕ₀, θ₀) are adimensional constants:

$$\mathbf{a}(\mathbf{j}_0,\mathbf{q}_0) = \frac{1}{4\mathbf{p}^4}\int\left[\int h(\mathbf{j}_i-\mathbf{j}_j,\mathbf{q}_i-\mathbf{q}_j)d\mathbf{j}_i\,d\mathbf{q}_i\right]d\mathbf{j}_j\,d\mathbf{q}_j$$

(26)

$$\mathbf{b}(\mathbf{j}_0,\mathbf{q}_0) = \frac{1}{4\mathbf{p}^4}\int\left[\int g(\mathbf{j}_i-\mathbf{j}_j,\mathbf{q}_i-\mathbf{q}_j)d\mathbf{j}_i\,d\mathbf{q}_i\right]d\mathbf{j}_j\,d\mathbf{q}_j$$

We nominate Φ as combined potential.

### 3.3 The Picture of Static Gravity and Electrostatics in Temporal Fluctuations Model

We focus now on charged particles and how they interact, gravitationally and electric, in the static case. Remembering that we show that, in TFM, a particle can be seen as a spherical region, which border's density of intensity is 2i₀ and the only difference between particles is due to their radius.

For a charged particle we must consider the asymmetry of fluctuation's distribution, orientated in such manner for explain interaction between charges. The simplest and convenient mode that can explain charge interactions is to consider radial asymmetry. For one kind of charge the asymmetry is oriented from charge to outside and for the other kind is oriented inversely. We don't have at this moment a criterion for consider which asymmetry is linked with positive charge and which is linked with negative charge. We'll see later that this picture can be linked with CPT symmetry. The asymmetry of fluctuation's distribution overlaps with density's intensity variation, so theirs effects combine.

We presume that the asymmetry starts from a small spherical region (with radius $r_e$) and propagate towards outside. At $r_e$ the asymmetry is maximum, i.e. $(\Delta n/n_0)=1$. We presume that the asymmetry diminished with distance r following the law:

$$\frac{\Delta n}{n_0} = \frac{r_e^2}{r^2} \qquad (27)$$

An argument for relation (27) is given in Appendix. It results that, at particle's border, the asymmetry is different from one kind of particle to other, because their radius differs.

Let's consider now two particles with different masses and charges with opposite sign (i.e. opposite asymmetry of fluctuation's distribution) - see figure 1. The maximum asymmetry is on direction that links the particles.

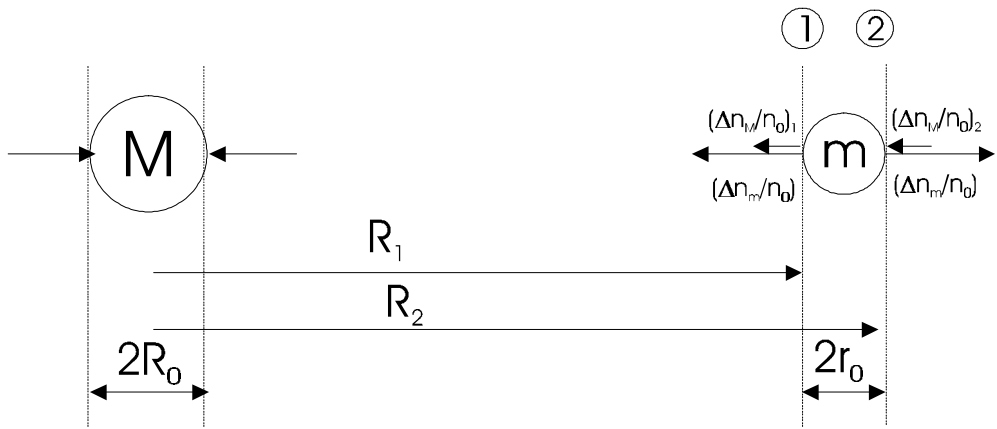

Figure 1

Using notation from figure, the combined potentials at borders 1 and 2 of test particle are:

$$\Phi_1 = C \cdot \Delta t \cdot (i_m + i_1) \cdot \left\{ \mathbf{a}(\mathbf{j}_0, \mathbf{q}_0) + \mathbf{b}(\mathbf{j}_0, \mathbf{q}_0) \cdot \left[ \left( \frac{\Delta n_m}{n_0} \right) + \left( \frac{\Delta n_M}{n_0} \right)_1 \right] \right\} \tag{28}$$

$$\Phi_2 = C \cdot \Delta t \cdot (i_m + i_2) \cdot \left\{ \mathbf{a}(\mathbf{j}_0, \mathbf{q}_0) + \mathbf{b}(\mathbf{j}_0, \mathbf{q}_0) \cdot \left[ \left( \frac{\Delta n_m}{n_0} \right) - \left( \frac{\Delta n_M}{n_0} \right)_2 \right] \right\}$$

Where:

- $i_m$ is density of intensity produce by particle m at its border;
- $i_1$ and $i_2$ are density of intensity produce by particle M at borders 1 and 2 of m.

We calculate the quantity:

$$\frac{\Phi_1 - \Phi_2}{2r_0} = C \cdot \Delta t \cdot \frac{(i_1 - i_2)}{2r_0} \cdot \left\{ \mathbf{a}(\mathbf{j}_0, \mathbf{q}_0) + \mathbf{b}(\mathbf{j}_0, \mathbf{q}_0) \cdot \left( \frac{\Delta n_m}{n_0} \right) \right\} + \frac{2 \cdot \mathbf{b}(\mathbf{j}_0, \mathbf{q}_0) \cdot i_m}{2r_0} \left( 1 + \frac{i}{i_m} \right) \left( \frac{\Delta n_M}{n_0} \right) \tag{29}$$

Where for R>>r$_0$ we can approximate:

$$\left( \frac{\Delta n_M}{n_0} \right) \cong \left( \frac{\Delta n_M}{n_0} \right)_1 \cong \left( \frac{\Delta n_M}{n_0} \right)_2 \quad \text{and} \quad i \cong i_1 \cong i_2 \tag{30}$$

From (10), for great distances from M we can write:

$$i \cong i_0 \left( 1 + \frac{GM}{c^2 R} \right) \quad \text{with} \quad R \cong R_1 \cong R_2 \tag{31}$$

Using the link between mass and radius, equation (29) become:

$$\frac{\Phi_1 - \Phi_2}{2r_0} = \frac{C \cdot \Delta t \cdot i_0}{c^2} \cdot \frac{GM}{R^2} \cdot \left\{ \mathbf{a} + \mathbf{b} \cdot \left( \frac{r_e}{r_0} \right)^2 \right\} + \frac{C \cdot \Delta t \cdot i_0}{c^2} \cdot \frac{3 \cdot \mathbf{b}(\mathbf{j}_0, \mathbf{q}_0) \cdot c^4 \cdot r_e^2}{G} \cdot \frac{1}{m_0 R^2} \left( 1 + \frac{1}{3} \cdot \frac{GM}{c^2 R} \right) \tag{32}$$

If we consider that:

$$\left| \frac{e_M e_m}{4\pi \varepsilon_0} \right| = \frac{3 \cdot C \cdot \Delta t \cdot i_0 \cdot \mathbf{b}(\mathbf{j}_0, \mathbf{q}_0) \cdot c^2 \cdot r_e^2}{G} \tag{33}$$

And:

$$C \cdot \Delta t \cdot i_0 \cdot \left\{ \mathbf{a}(\mathbf{j}_0, \mathbf{q}_0) + \mathbf{b}(\mathbf{j}_0, \mathbf{q}_0) \left(\frac{r_e}{r_0}\right)^2 \right\} = c^2 \qquad (34)$$

Equation (32) become:

$$\frac{\Phi_1 - \Phi_2}{2r_0} = \frac{GM}{R^2} - \frac{e_M e_m}{4\pi\varepsilon_0} \cdot \frac{1}{m_0 R^2} \left(1 + \frac{1}{3} \cdot \frac{GM}{c^2 R}\right) \qquad (35)$$

We must observe that, in absence of a more detailed model of fluctuation's interaction (who needs an experimental investigation), the quantities α, β and $r_e$ aren't determined.

In the first approximation, the parenthesis from equation (35) is equal to 1. In this situation, we observe that the right side of equation (35) is the radial acceleration (with changed sign) acquired by particle m in static fields of particle M. We observe too that quantity $\Phi_1 - \Phi_2$ is linked to the force that acts on test particle from source particle.

We can conclude that particle is moving under the difference of combined potential Φ between the opposite sides of particle and that it moves to region where Φ increase. We selected the sign of electrostatic part of acceleration to match the correct movement, i.e. in direction of increasing Φ.

We can presume that acceleration is intensity of a field for that we can define a potential Ψ (unified potential):

$$\vec{a} = -\operatorname{grad} \Psi \qquad (36)$$

We could write a similar relation between **a** and Φ only for the gravitational part of Φ (i.e. for dependence on i). For electrostatic part, we couldn't write such a relation because here we have both interacting particles. We can observe too that for Φ, the gradient have no senses for distances smaller than particle diameter.

On observe that we obtain naturally the independence from mass of acceleration due to gradient of density of intensity. On observe too the dependence from mass of acceleration due to gradient of asymmetry of fluctuations distribution.

This is easy to explain because the gradient of i depend on size of particle and the asymmetry of fluctuations distribution at opposite borders of particle do not depend. So, we see that inertia arise naturally from TFM.

Returning to equation (35), we observe the slight asymmetry of electrostatic interaction: the proton attract the electron a little bit stronger than electron attract the proton; this signify that neutral matter isn't quit so neutral. Could this explain the recent astrophysical conclusions on expanding Universe? Anyway, the asymmetries of electrostatic interactions produce effects that could be subject of experimental verification. Is this asymmetry the long time searched "the fifth force"?

### 3.4    Time Intervals

Equation (25) can be writing:

$$\Phi = C\alpha\Delta\tau \tag{37}$$

Where:

$$\Delta t = \Delta t \cdot i \cdot \left(1 + \frac{b}{a} \cdot \left(\frac{\Delta n}{n_0}\right)\right) \tag{38}$$

Is the average lifetime of fluctuations that depend on i and on distribution's asymmetry of fluctuations.

If we consider only the gravitational interaction (i.e. there is no asymmetry), we can write for lifetime of fluctuations in a region with increased i:

$$\Delta t = \Delta t \cdot i$$

And:

$$\Delta t_0 = \Delta t \cdot i_0$$

For vacuum.

From the two above equations we can write:

$$\frac{\Delta t}{i} = \frac{\Delta t_0}{i_0} \tag{39}$$

And with (10), we can write:

$$\Delta t = \Delta t_0 \cdot \left(1 + \frac{GM}{c^2 R}\right) = k \cdot \Delta t_0 \qquad (40)$$

It is obvious that in TFM the duration of all events depends on fluctuation's lifetime, so the same relation could be write with durations:

$$\Delta T = \Delta T_0 \cdot \left(1 + \frac{GM}{c^2 R}\right) = k \cdot \Delta T_0 \qquad (41)$$

The same relation as (41) result by considering that the change in energy produce a change of frequency processes, by $E = h\nu$. Result the following relation:

$$\nu = \nu_0/k \qquad (42)$$

From reciprocal of (42) we find the relation (41) between time intervals of such processes.

Equation (37) shows that combined potential and average fluctuation's lifetime are closely linked. The statement that particle is moving under gradient of combined potential is similar to the statement that particle is moving under gradient of fluctuation's lifetime or, more generally, that particle is moving under gradient of event duration.

### 3.5 The Metric Tensor

We had seen above that the factor k play main role for properties description of particles in gravitational field.

At this point, we can make a connection with metric tensor concept used in General Relativity formulation. We consider that far away from gravitational sources the vacuum is a 4-dimensional flat space. Relation gives infinitesimal interval:

$$ds^2 = c^2 dt_0^2 - (dr_0^2) \qquad (43)$$

For the same observer, the infinitesimal interval for a region in gravitational field is [16]:

$$ds^2 = \frac{1}{k^2} c^2 dt_0^2 - k^2 \left(dr_0^2\right) \tag{44}$$

Where on identify:

$$g_{00}=1/k^2, \qquad g_{11}=g_{22}=g_{33}=k^2, \qquad g_{ij}=0 \text{ for } i \neq j \tag{45}$$

For weak gravitational field we retrieve tensor elements for standard Schwarzschild metric:

$$g_{00} \cong \left(1 - \frac{2GM}{c^2 R}\right) \qquad g_{ii} \cong \left(1 + \frac{2GM}{c^2 R}\right) \tag{46}$$

As result, we can consider the movement in gravitational field as movement in gradient of average intensity's density.

We can extend k for gravitational and electrostatic fields. From (35) we write the radial part of acceleration:

$$a(R) = -\frac{GM}{R^2} + \frac{e_M e_m}{4\pi\varepsilon_0} \cdot \frac{1}{m_0 R^2}\left(1 + \frac{1}{3} \cdot \frac{GM}{c^2 R}\right) \tag{47}$$

By integration on radial part, we obtain from (36) the unified potential:

$$\Psi(R) = -\frac{GM}{R} + \frac{e_M e_m}{4\pi\varepsilon_0} \cdot \frac{1}{m_0 R}\left(1 + \frac{1}{6} \cdot \frac{GM}{c^2 R}\right) \tag{48}$$

Because for gravitational case k is:

$$k = 1 - \frac{\Phi_{gravitational}}{c^2} \tag{49}$$

We can obtain k for both fields considering $\Psi$ instead of $\Phi$. On write:

$$k = 1 + \frac{GM}{c^2 R} - \frac{e_M e_m}{4\pi\varepsilon_0} \cdot \frac{1}{m_0 c^2 R}\left(1 + \frac{1}{6} \cdot \frac{GM}{c^2 R}\right) \tag{50}$$

For the first approximation, relation (50) can be writes:

$$k = 1 + \frac{GM}{c^2 R} - \frac{e_M e_m}{4\pi\varepsilon_0} \cdot \frac{1}{m_0 c^2 R} \tag{51}$$

Metric tensor elements could be writing:

$$g_{00} \cong \left[1 - 2\cdot\left(\frac{GM}{c^2 R} - \frac{e_M e_m}{4\pi\varepsilon_0 m_0 c^2 R}\right)\right] \qquad g_{ii} \cong \left[1 + 2\cdot\left(\frac{GM}{c^2 R} - \frac{e_M e_m}{4\pi\varepsilon_0 m_0 c^2 R}\right)\right] \qquad (52)$$

We can observe that, apart from purely gravitational case when always k>1, for gravitational and electrostatic case we could have the situation k<1. The possibility of manipulation the value of k by means of electromagnetism could lead to very important technological applications, especially for transportation.

### 4. Lagrangean Approach for Static Fields

Given a particle with mass M and charge $e_M$ as source for gravitational and electrostatic fields and a test particle with rest mass $m_0$ (measured outside fields) and charge $e_m$. It is natural to consider that, as for gravitational field only, if test particle is approaching to source its mass become:

$$m = \frac{m_0}{k} \qquad (53)$$

We introduce (53) in Lagrange equation for a free particle:

$$\frac{d}{dt}\left(\frac{m\vec{v}}{\sqrt{1-\frac{v^2}{c^2}}}\right) = \nabla\left(-mc^2\sqrt{1-\frac{v^2}{c^2}}\right) \qquad (54)$$

Working with spherical symmetry, radial movements, stationary case and v<<c, result:

$$\frac{dk}{dR} = \frac{1}{c^2} a(R) \qquad (55)$$

Where:

$$a(R) = -\left( \frac{GM}{R^2} - \frac{e_M e_m}{4\pi \varepsilon_0 m_0 R^2} \right) \tag{56}$$

Is radial acceleration of test particle due to both fields. By integrating equation (55) with the same condition as for equation (5), it result:

$$k = e^{\frac{GM}{c^2 R} - \frac{e_M e_m}{4\pi \varepsilon_0 m_0 c^2 R}} \tag{57}$$

For weak fields, metric tensor elements could be write:

$$g_{00} \cong \left[ 1 - 2 \cdot \left( \frac{GM}{c^2 R} - \frac{e_M e_m}{4\pi \varepsilon_0 m_0 c^2 R} \right) \right] \qquad g_{ii} \cong \left[ 1 + 2 \cdot \left( \frac{GM}{c^2 R} - \frac{e_M e_m}{4\pi \varepsilon_0 m_0 c^2 R} \right) \right] \tag{58}$$

In the first approximation, from relation (57) we can write:

$$k = 1 + \frac{GM}{c^2 R} - \frac{e_M e_m}{4\pi \varepsilon_0} \cdot \frac{1}{m_0 c^2 R} \tag{59}$$

This is identical to (51) who was obtained in the frame of Temporal Fluctuations Model.

We can observe that if we substitute (57) in expression of free particle's Lagrangean, result the Lagrangean for the particle in weak static gravitational and electrostatic fields:

$$L = -\frac{m_0 c^2}{k} \cong -m_0 c^2 + \frac{GM m_0}{R} - \frac{e_M e_m}{4\pi \varepsilon_0} \cdot \frac{1}{R} = -m_0 c^2 - m_0 \Phi_{gravitational} - e_m \Phi_{electrostatics} \tag{60}$$

We conjecture that $m_0 c^2$ is particle's potential energy vis a vis of temporal fluctuations background (primary potential energy) and $\Theta = c^2$ is primary potential. If we consider energy conservation, we can conclude that all asymmetries influence the primary potential energy and, of course, the primary potential. These problems need a future development in a separate paper.

We conjecture too, that c from energy calculation is the velocity of fluctuations, therefore we can conclude from (1) that mass modify when enters in region with fields. The velocity of light, calculated from (44) by doing ds=0 :

$$c = \frac{c_0}{k^2} \cong c_0 \cdot \left( 1 + 2 \frac{\Psi}{c_0^2} \right) \tag{61}$$

Where $c_0$ is light velocity outside fields. Light velocity is only locally invariant, i.e. if it's locally measured.

Because k calculated in TF Model and with Lagrangean matches only in first order approximation, we conclude that TFM must need future improvements. That's why, from now on, we use for k relation (59). We can improve (59) by considering $m=m_0/k$ instead of $m_0$. Results:

$$k = \frac{1 + \dfrac{GM}{c^2 R}}{1 + \dfrac{e_M e_m}{4\pi\varepsilon_0} \cdot \dfrac{1}{m_0 c^2 R}} \tag{62}$$

With suitable approximation, (62) can be writes:

$$k \cong 1 + \frac{GM}{c^2 R} - \frac{e_M e_m}{4\pi\varepsilon_0} \cdot \frac{1}{m_0 c^2 R} - \frac{e_M e_m}{4\pi\varepsilon_0} \cdot \frac{GM}{m_0 c^4 R^2} + \left(\frac{e_M e_m}{4\pi\varepsilon_0} \cdot \frac{1}{m_0 c^2 R}\right)^2 + \left(\frac{e_M e_m}{4\pi\varepsilon_0}\right)^2 \cdot \frac{GM}{m_0^2 c^6 R^3} \tag{63}$$

We observe that appear terms that coupling both gravitation and electrostatics. In this situation, the unified potential can be writes:

$$\Psi(R) = -\frac{GM}{R}\left[1 - \left(\frac{e_M e_m}{4\pi\varepsilon_0} \cdot \frac{1}{m_0 c^2 R}\right)^2\right] + \frac{e_M e_m}{4\pi\varepsilon_0} \cdot \frac{1}{m_0 R}\left(1 + \frac{GM}{c^2 R} - \frac{e_M e_m}{4\pi\varepsilon_0} \cdot \frac{1}{m_0 c^2 R}\right) \tag{64}$$

If we deal with primary potential, we can conclude that the total potential at test particle's position can be writes:

$$\Theta - \Psi(R) = c^2 + \frac{GM}{R}\left[1 - \left(\frac{e_M e_m}{4\pi\varepsilon_0} \cdot \frac{1}{m_0 c^2 R}\right)^2\right] - \frac{e_M e_m}{4\pi\varepsilon_0} \cdot \frac{1}{m_0 R}\left(1 + \frac{GM}{c^2 R} - \frac{e_M e_m}{4\pi\varepsilon_0} \cdot \frac{1}{m_0 c^2 R}\right) \tag{65}$$

And k is the rapport between total potential $\Theta-\Psi(R)$ and primary potential $\Theta=c^2$:

$$k = \frac{\Theta - \Psi(R)}{\Theta} \tag{66}$$

We can understand this situation by considering that a particle, as a region with fluctuation's intensity $2i_0$ instead of $i_0$ for vacuum, have a potential energy (primary potential energy) vis a vis of vacuum. All asymmetries modify the primary potential and, in consequence, mass and primary potential energy (i.e. rest energy) modifies.

In case of neutral matter, we observe again that from equation (64) the electrostatic interaction is slightly asymmetric: the proton attracts the electron a little bit stronger than electron attracts the proton. The conclusions are the same as for equation (35).

## 5. Asymmetry of Fluctuation's Distribution and CPT Symmetry

We'll show that radial asymmetry of fluctuation's distribution can be relate to rotation of spatial axis around temporal one in quadridimensional space. That links directly the electrical charge with spatio-temporal symmetries in quadridimensional space and offers a support to deeper understanding of CPT symmetry.

We consider an event with coordinates $x_\alpha$, $\alpha=1,2,3,4$ ($x_4$ is temporal coordinate). During $dx_4$, the remaining three spatial axes rotate with infinitesimal angle $d\phi$, so $x_a$ (a=1,2,3) become $x_a+dx_a$. We can write:

$$x_\nu + dx_\nu = A_{\alpha\nu}x_\alpha \tag{67}$$

With conditions:

$$A_{\beta\alpha}A_{\nu\alpha}=\delta_{\beta\nu}; \qquad \delta_{11}=\delta_{22}=\delta_{33}=1 \quad \text{and} \quad \delta_{44}=-1 \tag{68}$$

Where:

$$A = \begin{bmatrix} \cos(df) & a_{12} & a_{13} & 0 \\ a_{21} & \cos(df) & a_{23} & 0 \\ a_{31} & a_{32} & \cos(df) & 0 \\ 0 & 0 & 0 & i \end{bmatrix} \tag{69}$$

The six unknown elements of matrix A can be found imposing conditions (68). Unfortunately, the resulting equations system can be solved only with approximation. For $d\phi$ very small, we obtain:

$$A \cong \begin{bmatrix} \cos(d\mathbf{f}) & 0 & 0 & 0 \\ 0 & \cos(d\mathbf{f}) & 0 & 0 \\ 0 & 0 & \cos(d\mathbf{f}) & 0 \\ 0 & 0 & 0 & i \end{bmatrix} \quad (70)$$

We obtain:

$$dx_a \cong -x_a \frac{d\mathbf{f}^2}{2} \qquad a = 1,2,3 \quad (71)$$

Which demonstrate that, in 3-D space, the resulting displacement is radial. For correct match with charge behavior (rotation in one direction lead to outside radial displacement and rotation in opposite direction lead to inside radial displacement), it should obtain $dx_a$ as function of first power of $d\phi$. Probably an exact solving of six equations system could lead to such a relationship.

### 6. Gravitodynamics and Electrodynamics

Let's consider the same particle with mass M and charge $e_M$ as source for gravitational and electrostatic fields and a test particle with rest mass $m_0$ (measured outside fields) and charge $e_m$. We renotate the unified potential given in (64) in the following way:

$$\Psi_0(R) = \frac{y}{R} \quad (72)$$

Where index 0 show that we deal with unified potential in case of static fields.

By analogy with electrodynamics, the unified potential in (x, y, z, t) of moving source particle with velocity v along Ox axis is:

$$\Psi(x,y,z,t) = \frac{y}{\sqrt{1-\frac{v^2}{c^2}} \cdot \left[ \left( \frac{x - v \cdot t}{\sqrt{1 - v^2/c^2}} \right)^2 + y^2 + z^2 \right]^{1/2}} \quad (73)$$

We can define the unified vector potential:

$$\vec{A}_\Psi(x,y,z,t) = \frac{\vec{v}}{c^2} \cdot \Psi(x,y,z,t) \quad (74)$$

In addition, unified quadrivector potential with components:

$$A_m = \left[\frac{1}{c}\Psi, \vec{A}_\Psi\right] \qquad m = 0,1,2,3 \tag{75}$$

If source particle moves with velocity **v**, the unified potential measured in the rest reference frame is:

$$\Psi' = \frac{\Psi - \vec{v}\cdot\vec{A}_\Psi}{\sqrt{1 - v^2/c^2}} \tag{76}$$

And k measured in the same reference frame is:

$$k' = \frac{\Theta - \Psi'}{\Theta} \tag{77}$$

By analogy with electromagnetism we define the intensity of the unified field:

$$\vec{E}_\Psi = -\nabla\Psi - \frac{\partial \vec{A}_\Psi}{\partial t} \tag{78}$$

And induction of field:

$$\vec{B}_\Psi = \nabla \times \vec{A}_\Psi \tag{79}$$

We can write the following equation for Lagrangean of test particle that moves with velocity **v** in gravitational and electromagnetic fields:

$$L = -\frac{m_0 c^2}{k'} \cdot \sqrt{1 - \frac{v^2}{c^2}} \tag{80}$$

For $\Psi' \ll \Theta$ equation (80) become:

$$L \cong -m_0 c^2 \cdot \sqrt{1-\frac{v^2}{c^2}} \cdot \left(1 + \frac{\Psi - \vec{v}\cdot\vec{A}_\Psi}{c^2 \cdot \sqrt{1-v^2/c^2}}\right) = -m_0 c^2 \cdot \sqrt{1-\frac{v^2}{c^2}} - m_0\Psi + m_0\vec{v}\cdot\vec{A}_\Psi \tag{81}$$

For $\Psi$ without terms who combine gravity with electromagnetism, the Lagrangean become:

$$L = -m_0 c^2 \cdot \sqrt{1-\frac{v^2}{c^2}} - m_0\Phi_{grav} - e_m\Phi_{el} + m_0\vec{v}\cdot\vec{A}_{grav} + e_m\vec{v}\cdot\vec{A}_{el} \tag{82}$$

That it is exactly the Lagrangean for a particle in electromagnetic and gravitational fields. We emphasize that appear the term for gravitomagnetic field.

Using Lagrange equation and definitions (78) and (79), we obtain equation of movements for test particle in gravitodynamic and electromagnetic fields:

$$\frac{d\vec{p}}{dt} = m_0 \vec{E}_\Psi + m_0 \vec{v} \times \vec{B}_\Psi \qquad (83)$$

Where in the right side of (83) it is the analogue of Lorentz force for unified field.

Detailing (83) we observe that appear Lorentz force and gravitational analogue of Lorentz force.

For complete form of Ψ, both Lagrangean and equation of movements become more complicated.

We can define also the "unified field charge density" - $\rho_\Psi$ and vacuum permitivity - $\varepsilon_\Psi$:

$$\frac{\rho_\Psi}{\varepsilon_\Psi} = -4\pi G \rho_M + \frac{\rho_{eM} \rho_{em}}{\varepsilon_0 \rho_m} \qquad (84)$$

Where $\rho_{eM}$ and $\rho_{em}$ are electric charge densities for source and for test particle. Also, $\rho_M$ and $\rho_m$ are mass densities for source and for test particle.

Formally, we can put $\rho_\Psi = \rho_M$ and the remaining expression of (84) for $\varepsilon_\Psi$. We observe that, in this situation, $\varepsilon_\Psi$ depend on specific charge of source and test particles. However, we consider that is better to remain at combined expression (84).

We define also the "unified field charge current density" by relation:

$$\frac{\vec{j}_\Psi}{\varepsilon_\Psi} = \vec{v} \cdot \frac{\rho_\Psi}{\varepsilon_\Psi} \qquad (85)$$

With above definitions, we can write the analogous of Maxwell equations:

$$\begin{aligned}
\nabla \cdot \vec{E}_\Psi &= \frac{\rho_\Psi}{\varepsilon_\Psi} \\
\nabla \cdot \vec{B}_\Psi &= 0 \\
\nabla \times \vec{E}_\Psi &= -\frac{\partial \vec{B}_\Psi}{\partial t} \\
c^2 \cdot \nabla \times \vec{B}_\Psi &= \frac{\vec{j}_\Psi}{\varepsilon_\Psi} + \frac{\partial \vec{E}_\Psi}{\partial t}
\end{aligned} \qquad (86)$$

With some calculations, we can obtain from (86) the Maxwell equations for electromagnetism and the analogous of Maxwell equations for gravitodynamics (see [6] and [18]).

Solving (86) with Lorentz like gauge on obtain the wave equations for $\Psi$ and $\mathbf{A}_\Psi$.

By analogy to electromagnetism, it can write the corresponding equations for unified field. Moreover, it can be use the quadridimensional form for writing the equations for unified field.

The complete development of unified field formalism will be the subject of a future paper.

## 7. Conjectures for a possible model of particles

Temporal Fluctuation Model can be the starting point to a new understanding of particles. Of course, up to date, we are far away to a coherent theory of particles based on TFM. However, we can make first step toward such a theory and only the experiments will validate or not such attempt.

We had seen that radius of particle determine its mass and that at particle's border the density of intensity is always $2i_0$ and it decrease to $i_0$ at infinite following a $1/r$ law. As well, we had seen that the only difference between particles is due to their radius. In addition, the asymmetry of fluctuation's distribution increasing the interactions between fluctuations and can explain electromagnetic interactions.

We conjecture that $2i_0$ it is minimum density of intensity for maintaining selforganisation, which determine that particle it is an entity. At particle's surface it is the border between inside selforganisation and outside extension of particle's properties that decreases with distance (i.e. unified field). The particle and their field are the same entity and the field is linked to particle. By field mediation, particles interact one with others. The changes in the state of particle movements determine the changes in field that propagate with finite velocity (perceived as electromagnetic and gravitational waves). The relative movements of two particles determine that the unified field perceived by one particle from the other one is altered vis a vis of static case and this is described by use of unified quadrivector potential instead of unified scalar potential. The relative movements: straight, circular and spin produce the same alteration of unified field and that is observed as appropriate magnetic field and gravitomagnetic field.

We conjecture also, that nuclear strong and weak interactions are due to nonlinear behavior of unified field near to particles. Probably those nonlinearities concur to particle stability.

Because we suppose that temporal fluctuations have a very short lifetime, we can't understand the particle as a "collection" of temporal fluctuations, but as a selforganised finite region with some properties that propagate as a wave (or, better, as a soliton). Beside particle, there is the outside particle's extension (the field – without selforgnisation capability) as extension of particle's properties that moves with particle and intermediate the interactions with others particles.

A coherent theory of particles must start with the deeper understanding of interactions between temporal fluctuations, including selforganisation. We conjecture that temporal fluctuations have a boson behavior, in sense that the probability to appear (in a given region) a fluctuation with given properties (intensity, lifetime and orientation) depend on number of fluctuations with same properties from that region. This fact can lead to explanations for selforganisation and wave propagation.

We can imagine that selforganisation process is merely dynamic instead of static process. In this way, we can treat particles as spherical oscillators. We could link this fact to the wave behavior of particles. The radius of stabile particles must form a discrete spectrum.

The theory must infer an equation for particle's radius (with discrete spectrum of solutions). Thus, we could obtain the spectrum of permissible radius of particles, which must be linked to asymmetry of fluctuation's distribution (charge) and spin.

Finally, we can speculate that the intensity of fluctuations is quantised with (at least) three values: 0, $I_0$ and $2I_0$. Value 0 is for non-existence of fluctuation, $I_0$ is the value of intensity for fluctuations of "pure" vacuum and $2I_0$ is the value of intensity for fluctuations from inside of the particles. For intermediate regions (i.e. fields), there is a mixture of fluctuations with $I_0$ and $2I_0$ intensities, the concentration of $2I_0$ fluctuation's intensity decreasing with distance.

The quantification of intensity leads to quantification of intensity's density i, because the number of fluctuations is a discrete quantity. This leads to quantification of mass, distance, duration, energy and finally, to action quantification – Planck's constant.

## 8.    Final Notes

Of course, the TFM is far away to be a complete theory. Still, the "blueprint" form presented here shows a unitary vision for gravity, inertia and electromagnetism. We obtain the main results of General Relativity and Electromagnetism and the theory, until now, don't contain self-contradictions. As well, we obtain the "unified potential" which has terms that coupling gravitation with electromagnetism and who could be the basis for future technological applications.

The theory is at very beginning and future developments will improve the theory and will reveal new implications. Also, the mathematical formulation of theory must be improved and we can imagine two ways for this: tensorial onset and fluid physics onset.

At present stage, we imagine three major technological applications: energy production, matter manipulation and space transportation. First two applications could derive from observation that, mainly, particles differ by their radius. We can imagine a technology that, based on a fields mixture, change the radius of particles with emission or absorption of energy. The third application could be possible by manipulation of intensity's density in different space regions.

### 9. Appendix

We can consider the quantity $\Delta n$ as flow of temporal fluctuations that enters or get out through surface unit in a determined duration. Because at $r_e$ the asymmetry is maximum, then there $\Delta n = n_0$ and the whole flow is $4\pi r_e^2 n_0$. At distance r the whole flow is $4\pi r^2 \Delta n$. Because the whole flow must conserve, we obtain:

$$\frac{\Delta n}{n_0} = \frac{r_e^2}{r^2} \tag{A1}$$

For density of average intensity, the radial component of density's gradient can be seen as a flow with $1/r^2$ relationship. We consider that radial component of density's gradient obey the following law:

$$\frac{di}{dr} = \frac{q'}{r^2} \tag{A2}$$

Where q' is constant. By integration with condition:

$$r \to \infty \quad \Rightarrow \quad i \to i_0$$

And with $-q' = q\, i_0$ we obtain:

$$i = i_0\left(1 + \frac{q}{r}\right) \tag{A3}$$